# Boosting the quality factor of Tamm structures to millions by quantum inspired classical annealer with factorization machine

Jiang Guo,[1] Koki Kitai,[1,2] Hideyuki Jippo,[3] and Junichiro Shiomi[1,2,4*]

[1]Department of Mechanical Engineering, The University of Tokyo, 7-3-1 Hongo, Bunkyo, Tokyo 113-8656, Japan

[2]Institute of Engineering Innovation, The University of Tokyo, 2-11-16 Yayoi, Bunkyo, Tokyo 113-8656, Japan

[3]Fujitsu Limited, 1-5 Omiyacho, Saiwai-ku, Kawasaki, Kanagawa 212-0014, Japan

[4]RIKEN Center for Advanced Intelligence Project, 1-4-1 Nihombashi, Chuo-ku, Tokyo 103-0027, Japan

*E-mail: shiomi@photon.t.u-tokyo.ac.jp

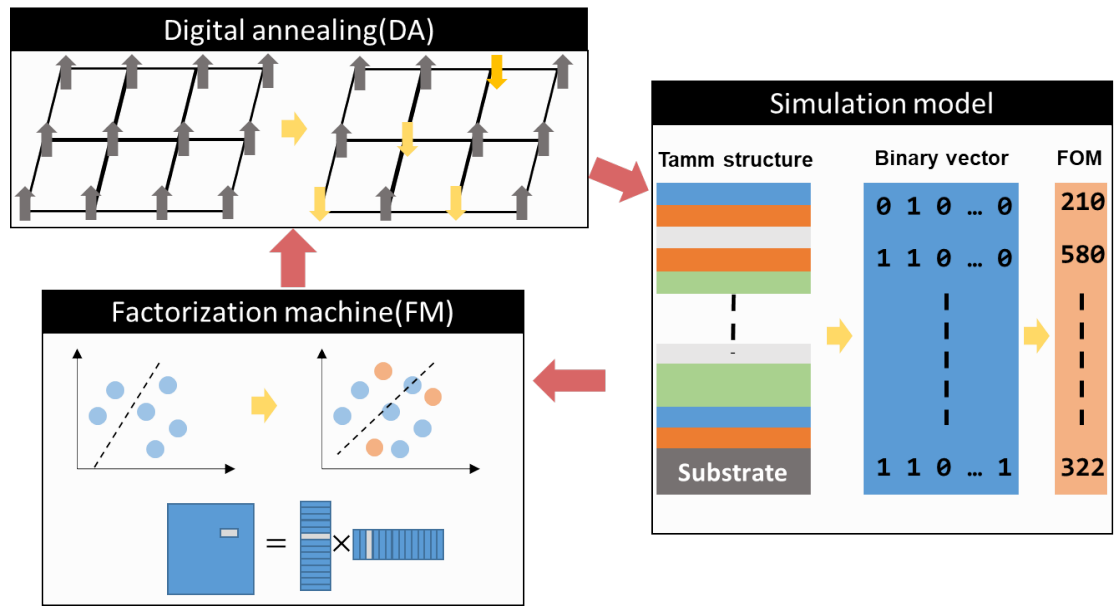

**Abstract**

The Tamm structures show high quality (Q) factor property with simple photonic Bragg reflectors composed of alternative high/low refractive index layers and metal reflectors. However, the Q-factor of Tamm structures is inherently limited by the periodic constraints and fixed thicknesses of the Bragg reflector pairs. Removing periodic constraints and adopting aperiodic designs can lead to significantly higher Q-factors. Nevertheless, to fully exploit the potential of Tamm structures, it is essential to thoroughly explore the extensive search space introduced by these aperiodic designs. Herein, we introduce a novel approach that utilizes the quantum-inspired classical annealer combined with a factorization machine to facilitate the design of photonic Bragg reflectors with the aim of achieving ultrahigh Q-factor properties. Our investigation begins by establishing the effectiveness of global optimization in a 15-bit problem setting, showcasing a remarkable efficiency improvement of nearly one order of magnitude compared to random search strategies. Subsequently, we expand our analysis to a 20-bit problem, demonstrating comparable efficiency to state-of-the-art Bayesian optimization methods. Moreover, the proposed method proves its capability in handling extremely large search spaces, exemplified by a 40-bit problem, where we successfully uncover a Tamm structure design exhibiting an extraordinary ultrahigh Q-factor surpassing millions. The designed aperiodic Tamm structure exhibits exceptional high localization and enhancement of electric field, and the power loss is concentrated in the low loss dielectric layer. Hence, the decay rate of power dissipation significantly decreases and light-matter interaction time increases, resulting in ultrahigh Q-factor.

# Introduction

Metamaterials, such as photonic crystals, surface-patterned structures, multilayer thin films, etc., have been proposed to regulate the light and matter interactions in spectral and spatial dimensions.[1–9] Among these artificial sub-wavelength structures, Tamm plasmon polariton (TPP) heterostructures have been fabricated using the cost-effective and thin film deposition method. Furthermore, they can support surface plasmon polariton (SPP) for transverse magnetic/electric (TM/TE) waves without an additional prism coupler setup because the surface plasmon dispersion line lies within the free-space light cone, ensuring that the surface plasmon excitation condition can always be satisfied.[7,10] Tamm state of TPP structures refers to the interface state between the metal mirror and periodic Bragg reflector mirror analogous to the electron state occurring at energy bandgap of crystal structures proposed by Tamm.[11,12] Tamm structures have various applications requiring strong confinement or trapping of light, such as nano-laser, gas detection, and narrowband thermal emitters.[8–10,13–15]

Various studies have focused on improving the quality (Q) factor of the Tamm structure, which describes the lifetime of the stored energy and bandwidth of the optical spectrum peak or dip. High Q Tamm structures improve the energy utilization efficiency in various applications, such as narrowband thermal emitter for thermophotovoltaic[8], ultra-selective and low-temperature-oriented radiative cooler[16], IR gas sensors[15], and radiator dryers[17] that absorb or radiate thermal radiation in a specific wavelength range. Yang et al. studied Tamm structures with different metal mirrors or Bragg-reflector pairs and compared the Q factors of metal side and Bragg reflector side structures.[10]

Symonds et al. reported the enhancement of light confinement by inserting dielectric spacer between the metal mirror and photonic reflector.[18] Zhu et al. proposed the bilayer-cavity enhanced Tamm structure, and achieved the Q-factor of approximately 170 with wavelength tunability varying the phase of the material.[19] Wang et al. proposed a combination of two Bragg reflectors with overlapping photonic bandgap to excite the so-called optical Tamm state (OTS), and experimentally determined a Q-factor approximately 1120 at 1.55 um.[20] Lin et al. used topology optimization to enhance the second harmonic generation.[21] Poddubny et al. analyzed the effect of perturbation in dielectric constant on the photonic bandgap reflection dips within Bragg reflectors and observed that sharp reflection dips can be induced by breaking the periodicity owing to Fano resonance.[22]

Previous studies indicates that periodic Bragg reflector design is not necessarily required for a high-Q Tamm structure, aperiodic Bragg reflectors are more flexible and need much less layers to realize strong light-structure interactions and high Q properties. While aperiodicity gives rise to vast structural candidates, an optimization methodology is crucial for designing Tamm structures with high-Q-factors. Christopher et al. used a genetic algorithm to optimize $SiO_2$/Si thickness for narrowband thermal emitters.[23] Sakurai et al. used Bayesian optimization (BO) to optimize the material choice and Bragg reflector thickness for narrowband thermal emitters, and obtained a high Q-factor of approximately 270 and 188 in the simulation and experiment, respectively.[17] So et al. constructed an artificial neural network to optimize the material choice and thickness of Tamm structures, wherein the simulated Q-factor of the optimal structure was

approximately 109.[24]

Tamm structure optimization is a standard combinatorial problem, which is a nondeterministic polynomial time (NP) hard problem.[25] Furthermore, the combinatorial problem can be mapped into the Ising model or quadratic unconstrained binary optimization (QUBO) to determine the ground state of energy levels.[26] Kitai et al. successfully applied the quantum annealer for high-efficiency radiative cooling metamaterial design.[27] The number of quantum annealers qubits has been gradually increasing; however, issues with scalability persist.[28] Hence, classical annealers attract attention owing to a large number of available bits. In this study, a second-generation Fujitsu Quantum-inspired Computing Digital Annealer (DA) was used to solve fully-connected QUBO problems. The DA is based on simulated annealing algorithm, which has the following distinctions: initial state, jumping-out energy barrier setting, parallel rejection algorithm.[29] The DA has been demonstrated to out-perform quantum annealer for specific problems with appropriate settings.[30]

**Results and Discussion**

In this study, the factorization machine model is used as the prediction model, which is combined with DA as the sampler (FMDA) to optimize the aperiodic Tamm structure for a high-Q photonic structure design. The small search space problem was addressed using FMDA to optimize a 15-layer Bragg reflector, thereby achieving approximately 10 times faster optimization efficiency compared to random sampling. Furthermore, the global optimal structure was determined, and FMDA was compared with BO for a medium-sized 20-layer Bragg reflector, demonstrating a comparable optimization efficiency. Finally, high-Q-factor designs were discovered using FMDA to optimize the material types and thicknesses in the 20-layer Bragg reflector Tamm structures.

Figure 1 shows the working procedure of the proposed FMDA for ultrahigh-Q Tamm structure design. The Tamm structures corresponding to the binary vectors were evaluated using transfer matrix method (TMM), an electromagnetic (EM) solver. The Q-factor and other defined figures of merit were calculated based on the EM simulation results. Initially, the surrogate FM model was trained by collecting sufficient data for pre-training. Thereafter, the trained FM model outputs the weight matrix and bias vectors, and these parameters are further embedded into a QUBO representation for sampling using the digital annealer. Notably, since the weight matrix from FM is decimal formal and upper diagonal, which is inconsistent with the DA annealing set, so these parameters will be converted to DA compatible format before annealing. The annealing process would output several binary vector outputs exhibiting close minimum energy level (current stage optimal solution). The binary vector and the corresponding figure of merit (FOM) value were added as new data points for the FM model training and update. Thereafter, the updated FM model could output new weight matrix and bias parameters for DA sampling. After finite iterations, the prediction accuracy of the FM model improves, and the ultrahigh-Q-factor Tamm structures can be optimized.

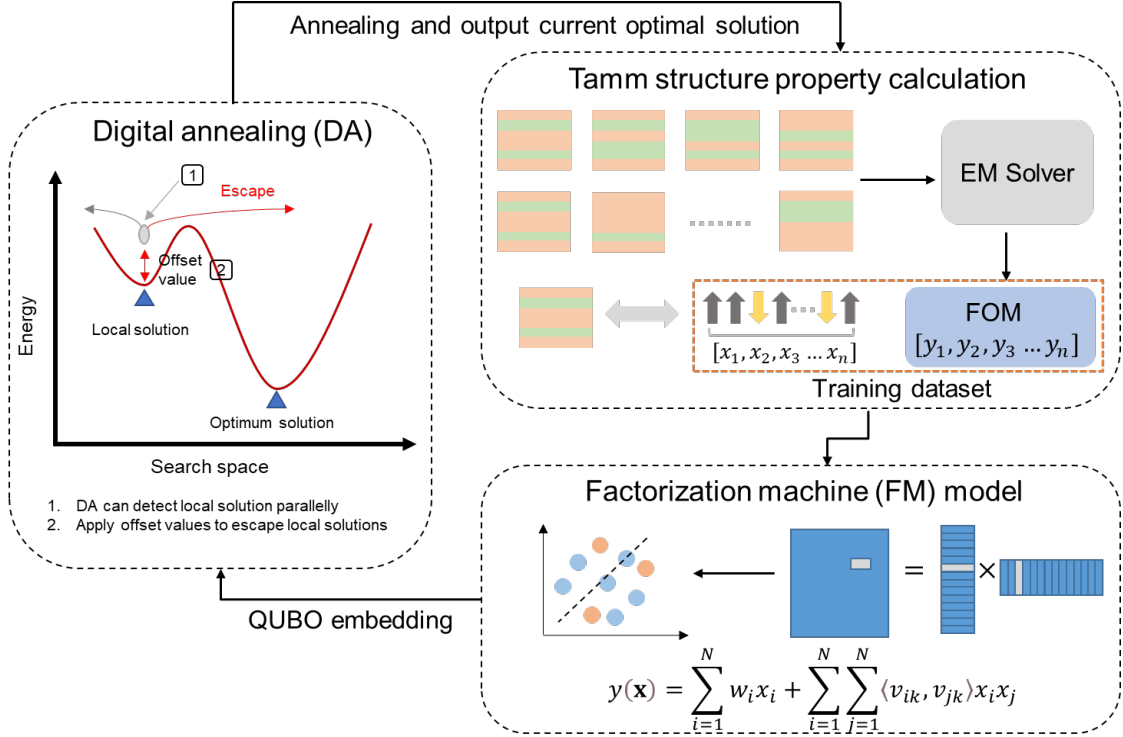

**Figure 1**. FMDA procedure for high-Q Tamm structure design

To confirm the global optimization property of FMDA, we optimized the photonic model of a 15-layer Tamm structure consisting of Ge and SiO$_2$ (labeled as '0' and '1') corresponding to the high and low refractive index material, respectively. Herein, the optically opaque tungsten (W) was used as the substrate, and the material optical constant data were obtained from the Palik database.[31] For trial and demonstration, the total thickness of the Tamm structure model was fixed to 9 $\mu m$, i.e., each unit layer was 600-nm thick. The wavelength range for EM simulation was in the range 5 to 6 $\mu m$. The wavelength spacing of the spectrum peak was automatically refined during the simulation to ensure that the number of valid data points for the peak was greater than ten. The figure of merit was set as the Q-factor. The binary vectors corresponding to the photonic structures were optimized using FMDA. Three independent runs of FMDA and random sampling optimization were performed, and the results were compared in Fig. 2(a). Note that, the FMDA results refer to the top axis labels and the random sampling optimization results refer to the bottom axis labels. Furthermore, all FMDA runs rapidly converged to the global optimum, and the optimization efficiency of FMDA was approximately ten times greater than that of random sampling optimization. Figure 2(b) shows the histogram plot of the high Q-factor of Tamm structures, confirming that its optimized value is the global optimal value. The Q-factors of all candidate Tamm structures were calculated; however, only those with Q-factor values greater than 3000 are shown for clarity. As can be seen from the histogram plot, only one structure within the total candidate space exhibits the highest Q-factor, which is identified using the optimization method.

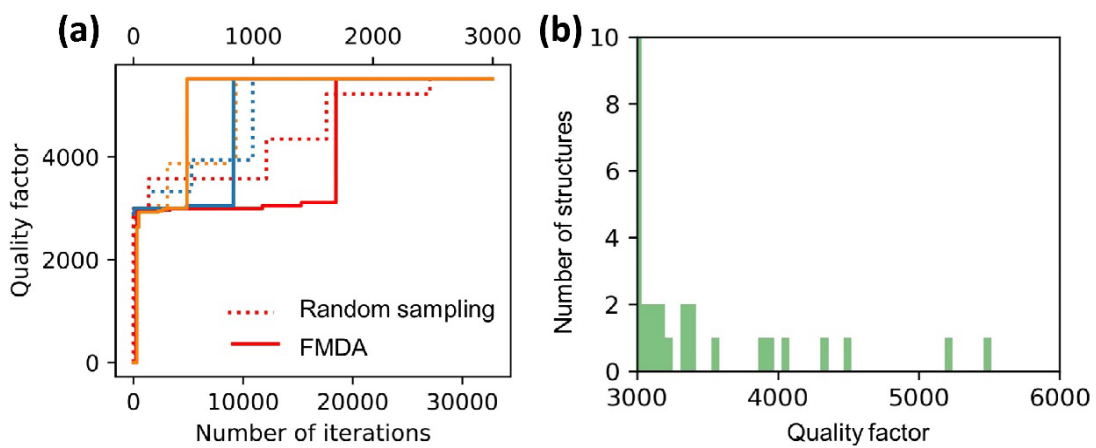

**Figure 2.** (a) Three independent runs of FMDA (solid lines) and random sampling optimization (dash lines) histories for the 15-layer Tamm structure optimization. The

FMDA and random sampling results refer to the top- and bottom-axis labels, respectively. (b) Histogram number plot of the optimized high-Q Tamm structures.

Thereafter, the optimization efficiency of BO was compared to that of the proposed FMDA method. The BO is highly efficient in determining the global optimum within a moderate candidate search space.[17,32–36] Herein, we adopted a Tamm structure model similar to that in the 15-bit optimization with a total thickness of 9 *µm*; however, the total number of layers was set to 20. The total number of candidates for this 20-bit optimization was $2^{20}$=1048579, which is suitable for fast BO. To design high-Q Tamm structures with sufficiently large absorption or emissivity, two figures of merit were defined and applied to the optimization. The first one (FOM-Q) is the Q-factor of the structure used when the magnitude of absorption or emissivity to be greater than 80%. The figure of merit was set to zero when the absorption or emissivity was less than 80%. The second figure of merit (FOM-A) is the product of the Q-factor and peak magnitude. To exclude the random effects of the initial training data, 20 independent runs of BO and FMDA were performed using FOM-Q and FOM-A. The averaged results of 20 independent runs were compared in Fig. 3(a). The FOM-Q optimization results were similar for BO and FMDA. For FOM-A, the overall averaged optimized results of FMDA were better than those of the BO. The difference in results between FOM-Q and FOM-A could be due to the fact that FOM-Q only considers the Q factor of the structure spectrum, while FOM-A accounts for both the Q factor the peak magnitude. Therefore, FMDA exhibits optimization efficiency competitive to BO for the 20-bits problem where the dimension of search space (or number of candidates) is small enough to perform global optimization. However, FMDA can be readily applied to larger dimensional search spaces because the sampling process is performed using an annealing machine, which does not increase the computational load as the dimensionality increases, unlike BO, which experiences an exponential increase in computational load with increasing dimensions.

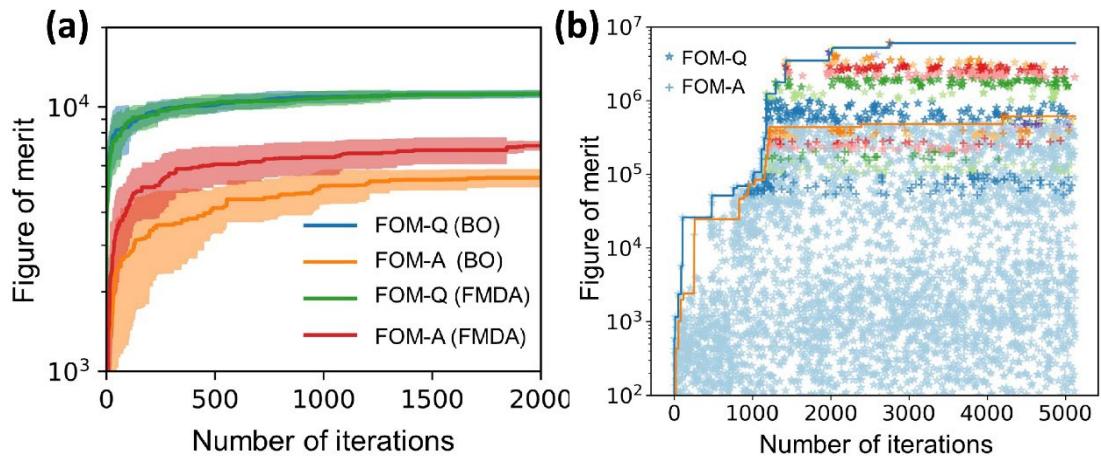

**Figure 3.** (a) The average of optimization results of the 20 runs of BO, which were compared to that of FMDA for 20-bit Tamm structure model. FOM-Q refers the figure of merit that is the quality factor when emissivity value is over 80% and set to 0 otherwise. FOM-A refers to the figure of merit that is the product value of Q-factor and absorption or emissivity magnitude. (b) FMDA optimization history results of the 40-bit Tamm structure model for FOM-Q and FOM-A.

Thereafter, Si and $MgF_2$ (optical constant data extracted from Palik book[31]) were included as new materials for the 20-layer Tamm structural model. The material vectors were labeled with two bits in the material choice list (labeled as '00' , '01', '10', '11'); thus, the total number of bits for this optimization was 40, which is approximately 1000 billion. The extremely high computational cost of BO in a high-dimensional space prohibits its application in the 40-bit optimization. Since the search space of the problem is extensive, identifying the best possible solution through numerous iterations is not practical. However, conducting the optimization process for a reasonable number of iterations has been shown to be effective in identifying structures with significantly improved performance. The proposed FMDA method can rapidly improve the figure of merit within several thousand iterations. Thus, to balance the computational time cost and figure of merit, the optimization iteration number was terminated at approximately 5000. The optimization results for FOM-Q and FOM-A are shown in Fig. 3(b).

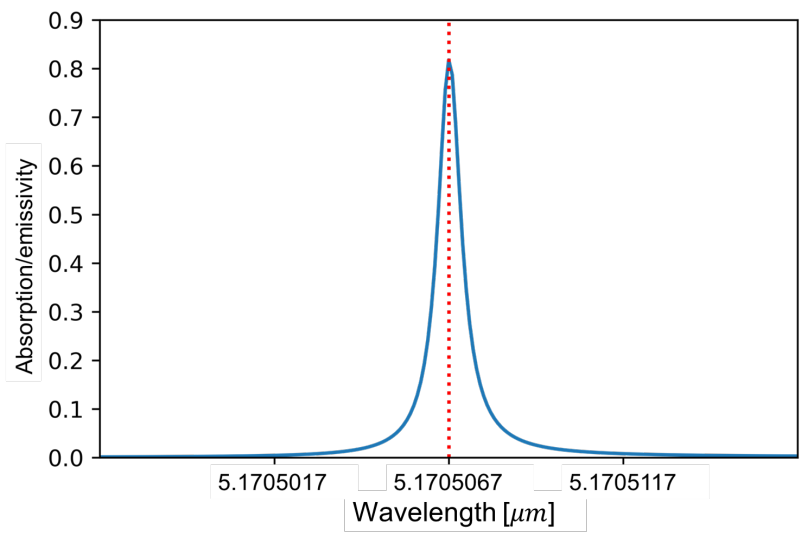

**Figure 4.** Absorption (A) plot of the optimized Tamm structure of the 4-material model, whose Q-factor is approximately 6 million and centered at 5.1705067 $\mu m$.

Notably, for FOM-Q and FOM-A optimizations, the Q-factors optimized by FMDA are far greater than those reported in the literatures.[7,9–11,13,14,17–19,23,24] The optimization of the 4-material Tamm structure resulted in a Q-factor approximately two orders of magnitude higher than the standard two-material structure. This indicates that incorporating material mixing in the Bragg reflector is advantageous for achieving enhanced Q factor. The absorption plot of the optimized 4-material Tamm structure is shown in Fig. 4, indicating an extremely narrow peak centered at wavelength of around 5.1705067 $\mu m$ with Q-factor greater than 6 million and absorption magnitude over 80%. Considering the thin film deposition precision, material impurities, and interfacial roughness, the optimized Tamm structure with an extremely high-Q-factor may be unrealistic for actual experimental fabrication. However, the aperiodic Tamm structure using rational optimization can achieve significantly stronger light-matter interactions and higher Q-factor compared to its periodic counterpart.

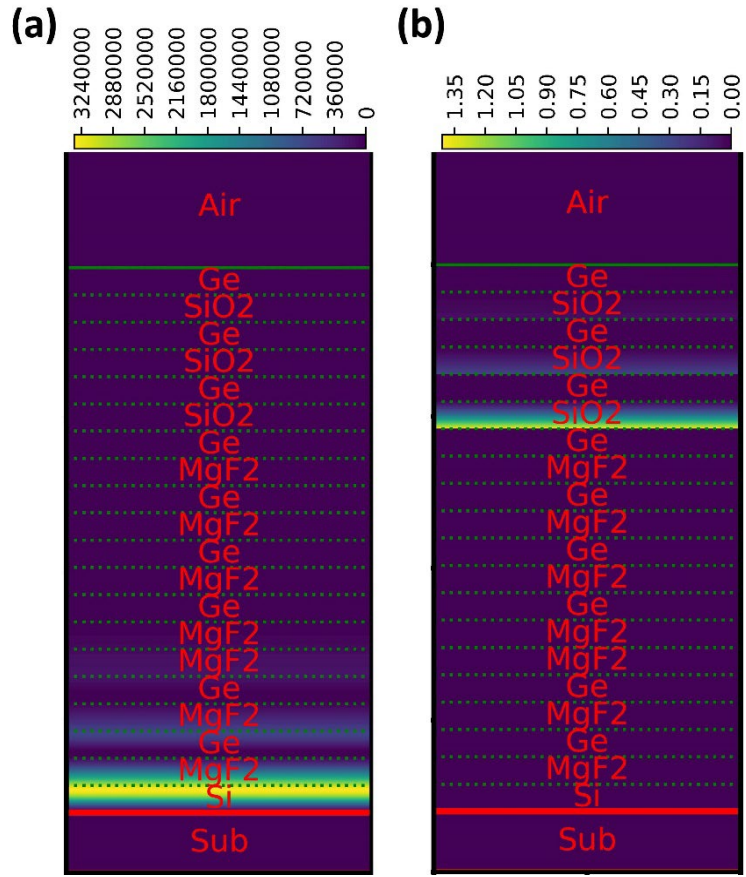

**Figure 5.** (a) Electric field and (b) power dissipation distributions of the optimized 4-material based ultrahigh-Q Tamm structure

Furthermore, the electromagnetic field distribution of the optimized 4-material ultrahigh-Q Tamm structure was meticulously examined. Fig. 5(a) illustrates the strong localization of the electric field at the interface between Si and MgF$_2$, exhibiting exceptionally high magnitudes. This observation signifies the remarkable enhancement of light-matter interaction within the optimized structure. The incorporation of aperiodic design in the Tamm structure introduces additional degrees of freedom. The disorder-induced localization confines the light within specific regions, causing a significant increase in electric field intensity and prolonging the duration of light-matter interaction. Consequently, the optimized structures exhibit a longer lifetime for the localized photons. This extended photon entrapment period allows for stronger interaction with the surrounding medium, thereby contributing to the ultrahigh Q-factor. Moreover, aperiodicity plays a crucial role in reducing radiative and non-radiative losses. The precise control over the photonic bandgap achieved by the non-repeating arrangement of layers in optimized aperiodic Tamm structures reduces the overlap between Tamm states and the continuum of radiation modes, thereby minimizing radiative losses. In contrast to periodic structures where the electric field peak is often localized near the high-loss metal substrate, the optimized aperiodic Tamm structure showcases a shift in the electric field peak away from the metal substrate. The power absorption in the optimized aperiodic Tamm structure predominantly occurs in the middle low-loss dielectric layer, specifically in the middle $SiO_2$ layer region. This displacement minimizes the interaction between light and high-loss regions, resulting in reduced non-radiative loss rate. This concentration of power absorption, combined with a low absorption rate within the operating wavelength range, prolongs light-matter interaction time, contributing to the ultrahigh-Q property of the structure.

In summary, the FMDA method was developed to optimize ultrahigh-Q Tamm structures. The capability and efficiency of the FMDA method in performing global optimization are demonstrated through the 20-bit problem, where its efficiency is comparable to that of BO and even better when adopting FOM-A. Unlike BO, which faces limitations in search space size due to the exponential growth of computational cost with dimensions, FMDA can be applied to larger search space problems. This is possible because the DA technique obtains the lowest energy solution through sampling, and the dimension of the problem is only limited by the sampler's bit number. Using FMDA, an ultrahigh-Q Tamm structure was successfully obtained with high efficiency for the 40-bit problem. The optimization of aperiodic Tamm structures through FMDA involves underlying physical mechanisms such as disorder-induced localization, reduced radiative losses, and power dissipation rate. These mechanisms enhance light-matter interaction and achieve an ultrahigh-Q factor.

**Methods**

Transfer matrix method

The Transfer Matrix Method (TMM) is a rigorous approach for simulating the optical properties of multilayer systems. It is particularly useful for modeling light propagation through stratified media, such as thin film layers or multilayered structures. The optical behavior of each layer is described by a 2x2 matrix, known as the transfer matrix, which relates the amplitudes of the forward and backward propagating waves at one interface to the amplitudes at the next interface. For an incident wave on a layer with a refractive index $n$ and thickness $d$, the transfer matrix $M$ of a single layer can be described as:

$$M = \begin{bmatrix} cos(\delta) & i(n/n_0)sin(\delta) \\ i(n_0/n)sin(\delta) & cos(\delta) \end{bmatrix} \quad (1)$$

where $\delta = 2\pi nd/\lambda$ is the phase change upon propagation through the layer, $n_0$ is the refractive index of the medium from which the light is incident, and $\lambda$ is the wavelength of light in vacuum.

For a multilayer system, the total transfer matrix $M_{total}$ is the product of the individual

layer matrices:

$$M_{total} = M_N M_{N-1} \ldots M_2 M_1 \tag{2}$$

where N is the total number of layers, and the multiplication is carried out from the substrate side towards the incident medium. The reflection and transmission coefficients for the multilayer system can be calculated from the elements of $M_{total}$. If $M_{total}$ is defined as:

$$M_{total} = \begin{pmatrix} a & b \\ c & d \end{pmatrix} \tag{3}$$

Then, the reflection and transmission coefficients, r and t, respectively, can be calculated by:

$$r = -c/a \tag{4}$$

$$t = 1/a \tag{5}$$

The optical results, such as reflectance R and transmittance T, can then be obtained from the square of the absolute values of r and t, respectively:

$$R = |r|^2 \tag{6}$$

$$T = |t|^2 * (n_{substrate}/n_{incident}) \tag{7}$$

Applying these principles, the TMM allows precise computation of the optical response of multilayer structures, accounting for interference effects and the individual properties of each layer. For complex structures, this method can be computationally intensive but provides accurate results.

Factorization machine

Factorization machine (FM) model[37] combines the support vector machines (SVM) and factorization models. Factorization model is regression model that defines the relationship between results and associated factors. Factorization machines using factorization models to reduce the correlation dimensions of interaction between variables, thus it is much easier to do prediction when correlation matrix between variables is extremely huge but sparse like the recommendation system where SVM fails to model the interaction between variables. The model for factorization machine for degree=2 can be written as,

$$y(\mathbf{x}) = \sum_{i=1}^{N} w_i x_i + \sum_{i=1}^{N}\sum_{j=1}^{N} \langle v_{ik}, v_{jk}\rangle x_i x_j, \qquad (8)$$

in which, $x_i$ is the variable to be modeled, $w_i$ models the strength of the $x_i$, $\langle v_{ik}, v_{jk}\rangle$ is the dot product of two variable vectors of size *K* (dimensionality of factorization),

$$\langle v_{ik}, v_{jk}\rangle = \sum_{k=1}^{K} v_{ik} v_{jk} \qquad (9)$$

This product models the interaction between $x_i, x_j$ variables by factoring the weight matrix, which allows the high-quality parameter estimation of high order interactions under sparsity.

Digital annealer

Digital annealing (DA) processor developed by the Fujitsu company aims to serve as the bridge that fill the gap between classic computer and the quantum computer. Inspired by the quantum annealing process, they have built the digital circuit to emulate the real quantum annealing process by simulated annealing with special offset technology to better jump out local energy minimal. In order to use the quantum inspired digital annealer which can efficiently acquire the lowest energy state, the quadratic unconstrained binary optimization (QUBO) is applied [38], which can be written as,

$$H = \sum_{i=1}^{N} \sum_{j=1}^{N} Q_{ij} x_i x_j, \tag{10}$$

where $x_i$ is the 0/1 binary bit and $Q_{ij} = Q_{ji}$ takes real values. Note that Q is upper-diagonal matrix which describes the interaction between variables. If the optimization variable number is very large, $Q$ will be usually sparse (lots of zero element) and difficult for machine learning algorithm to update the weight and fail to do prediction. Thus, FM regression model can be a good solution for large data QUBO problem.

The digital annealer searches the lowest energy level by binary quadratic formulation based on the Markov Chain Monte Carlo method. The energy function can be written as[29],

$$\boldsymbol{E}(\boldsymbol{x}) = -\sum_i \sum_{j \neq i} W_{i,j} x_i x_j - \sum_i w_i x_i \tag{11}$$

in which $x_i$ is the state value that represent the binary bit, $W_{i,j}$ is the coupling coefficient between bits, $w_i$ is the bias term. Digital annealer is supposed to have fully connected 8192 bits with high resolution of coupling coefficient. Thus, a large number of combinatorial problems can be mapped to the digital annealer system. The digital annealer algorithm is based on simulated annealing but with massive parallel operation. The annealing starts with the same arbitrary states, then it uses the parallel-trial search in which Monte Carlo consider a flip of state for individual variable. If there is one bit flip is accepted, one of the accepted bits will be selected uniformly. The state of all variables will be updated, and the field will be also changed in parallel. If there is no bit to update, the annealing will return to the trail state. The process will be repeated for several iterations until terminated by pre-set numbers. If the energy state falls to the local, narrow, and sharp energy dip of the energy function, it will be difficult for simulated annealing to escape the local optimal even with massive parallel search. The digital annealer is developed with a offset technology to jump out this local energy state. If there is no updated variable, the escape from the local minimum will be added by positive offset energy which is dynamically controlled to increase the possibility of finding the next flip bit and reduce the time trapped in the local minimal state. The digital annealer also implement the replica exchange method that independently and parallelly to perform the annealing process to accelerate the search for lowest energy level. When the lowest state is obtained by one trial and used as solution, the target state can be much easily to be found and the annealing time can be expected to be reduced. At the same time, thermal equilibrium state through Markov Chain Monte Carlo search can be also accelerated by this replica exchange method. The exchange probability between variables can be calculated by temperature an energy difference as indicated in the equation[30],

$$P_{n,n+1} = \min(1, \exp(\beta_n - \beta_{n+1})) * (E(X_n) - E(X_{n+1})) \qquad (12)$$

where n is the index of replica, $\beta_n$ is the inverse temperature, $X_n$ is the state of variables of replica n, $E(X_n)$ is the energy of replica n calculated by quadratic model. Through this replica exchange method, a path can be established between the higher temperature case and lower case, thus increase the probability to escape the local minimum energy state. The digital annealer has implemented both hardware based parallel search and independently replica exchange method to accelerate the annealing.

**Data availability**

All data needed to evaluate the conclusions in the paper are present in the paper. Additional data related to this study are available from the corresponding author upon reasonable request.

## Code availability

Optical reflection/transmission/absorption was simulated by home-made transfer matrix method python code. Factorization machine (FM) prediction model was from the open-sourced package FMQA[27]. The digital annealing sampler sampling code was adapted from Fujitsu company. All relevant codes are available from the corresponding author upon reasonable request. The code is provided to ensure reproducibility of the results presented in this paper and to facilitate further research.

## Acknowledgements

This work was partially supported by CREST Grant No. JPMJCR21O2 of the Japan Science and Technology Agency.

## Author information

Authors and Affiliations

**Department of Mechanical Engineering, The University of Tokyo, Tokyo, Japan**

Jiang Guo, Koki Kitai & Junichiro Shiomi

**Institute of Engineering Innovation, The University of Tokyo, Tokyo, Japan**

Koki Kitai & Junichiro Shiomi

**Fujitsu Limited, Kawasaki, Kanagawa, Japan**

Hideyuki Jippo

**RIKEN Center for Advanced Intelligence Project, Tokyo, Japan**

Junichiro Shiomi

## Contributions

J.S. conceived the idea and supervised the entire work. J.G. performed the optical simulation and FMDA optimization, analyzed the data and wrote the first draft of the paper. K.K. contributed to FMDA code framework. H.J. provide technical assistance for DA. All authors contributed to writing and editing the manuscript.

## Corresponding author

Correspondence to Junichiro Shiomi.

## **Ethics declarations**

Competing interests

The authors declare no competing interests.